\documentclass[
   superscriptaddress,
    prb,
    preprintnumbers,
   preprint,
    showpacs,
    fleqn,
    a4paper,
    showkeys
]{revtex4}

\usepackage{graphicx}
\usepackage{amsmath}
\usepackage{amsfonts}

\newcommand{\DifrOp}[1]{\left[\left[#1\right]\right]}

\newcommand{\dv}{\mathop{\rm div}\nolimits}
\newcommand{\1}[1]{#1^{(1)}}
\newcommand{\2}[1]{#1^{(2)}}

\begin{document}

\title{Pressure on charged domain walls and additional imprint mechanism in ferroelectrics}

\author{P. \surname{Mokr\'{y}}}
\email{pavel.mokry@tul.cz}%
\affiliation{Department of Electrical Engineering, Technical
University of Liberec, CZ-46117 Liberec, Czech Republic}

\author{A. K. \surname{Tagantsev}}
\affiliation{Ceramics Laboratory, Materials Department, EPFL Swiss
Federal Institute of Technology, CH-1015 Lausanne, Switzerland}
\date{\today}

\author{J. \surname{Fousek}}
\affiliation{Department of Electrical Engineering, Technical
University of Liberec, CZ-46117 Liberec, Czech Republic}

\date{\today}

\begin{abstract} The impact of free charges on the local pressure on a charged 
ferroelectric domain wall produced by an electric field has been analyzed. A 
general formula for the local pressure on a charged domain wall is derived 
considering full or partial compensation of bound polarization charges by free 
charges. It is shown that the compensation can lead to a very strong reduction 
of the pressure imposed on the wall from the electric field. In some cases this 
pressure can be governed by small nonlinear effects. It is concluded that the 
free charge compensation of bound polarization charges can lead to substantial 
reduction of the domain wall mobility even in the case when the mobility of 
free charge carriers is high. This mobility reduction gives rise to an 
additional imprint mechanism which may play essential role in switching 
properties of ferroelectric materials. The effect of the pressure reduction on 
the compensated charged domain walls is illustrated for the case of 
180${}^\circ$ ferroelectric domain walls and of 90${}^\circ$ ferroelectric 
domain walls with the head-to-head configuration of the spontaneous 
polarization vectors.
\end{abstract}



\maketitle

\section{Introduction}
\label{sec:Intro}

Understanding the dynamics of domain wall motion is essential for the 
explanation of many phenomena in ferroelectric materials. In many models for 
the evolution of domains in ferroelectrics, researchers deal with systems where 
charged ferroelectric domain walls appear. Classical examples to be mentioned 
are the work by Landauer \cite{Landauer} on the nucleation of 180${}^\circ$ 
domains and that by Miller and Weinreich\cite{Miller} on the sidewise movement 
of 180${}^\circ$ domain walls. In these models, the domain wall dynamics is 
analyzed by minimizing the thermodynamic potential consisting of the 
depolarizing field energy, the energy associated with the crystal lattice 
polarization, the energy of the domain wall, and the energy supplied to the 
system by the external electric source. It is evident that in all systems where 
charged domain walls appear there exists a possibility of compensation of bound 
charges by free charges, which affects the aforementioned energies through the 
reduction of the net charge at the domain wall. It was already pointed out by 
Landauer \cite{Landauer} that the process of bound charges compensation will 
not only reduce the depolarization energy, but it will also reduce the energy 
supplied to the system by external electric sources. Since the latter energy 
actually represents the driving force for domain wall motion, one can expect 
that the appearance of free charges in the system will result in a reduction of 
local pressure on the domain wall. This may have a serious impact on the domain 
nucleation and the sidewise domain wall motion. The problem of the impact of 
bound charge compensation on the domain pattern evolution seems to be of 
special practical interest for ferroelectric ceramics where the appearance of 
charged domain walls is readily expected whereas the high temperatures used to 
pole it can promote free-charge screening effects.

The aforementioned issues have motivated the theoretical analysis presented 
below, where we will address the effect of bound charge compensation on the 
local pressure exerted on the walls by the electric field. We will cover the 
general situation as well as some cases of special interest. In Sec. 
\ref{sec:Concept} we will illustrate the effect of charge compensation for the 
simplest situation. Section \ref{sec:Generalized} presents a derivation of the 
general expression for the local pressure imposed by the electric field to an 
arbitrarily compensated ferroelectric wall. In Sec. \ref{sec:PressureSpecial} 
we will demonstrate the application of the general results to systems with low 
electric fields (Subsec. \ref{sec:sub:LowFields}) and to systems where 
nonlinearity of the dielectric response of the crystal lattice is essential  
(Subsec. \ref{sec:sub:NonlinearEffects}).

\section{Simple model with full charge compensation}
\label{sec:Concept}

\begin{figure}
    \includegraphics[width=85mm]{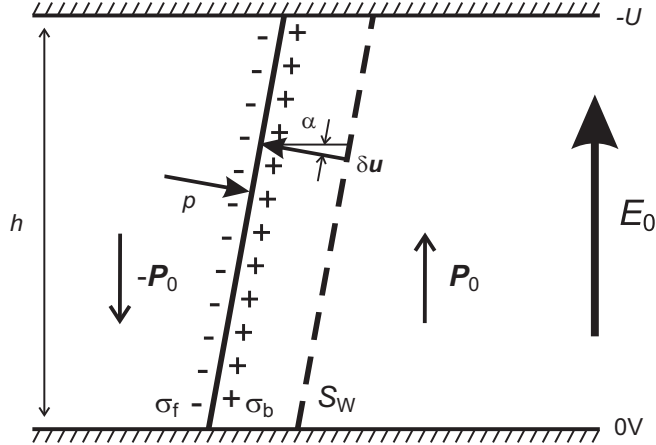}
	\caption{Scheme of a ferroelectric capacitor of thickness $h$ with a 
180${}^\circ$ ferroelectric domain wall tilted by an angle $\alpha$ with 
respect to its electro-neutral orientation. Bottom electrode is grounded and 
top electrode is at a constant electric voltage $-U$. Pluses indicate the 
positive bound charges $\sigma_b$ due to a discontinuity of the spontaneous 
polarization $P_0$ at the domain wall, minuses indicate the free charges 
$\sigma_f$ that fully compensate the bound charges. Symbol $\delta u$ stands 
for the virtual displacement of the domain wall and symbol $p$ stands for the 
external mechanical pressure that should be applied to the wall to keep it at 
rest.}
    \label{fig01a}
\end{figure}
At first we analyze the effect of free charges on the local pressure on a 
ferroelectric domain wall for the system shown in Fig. \ref{fig01a}. We 
consider the hard ferroelectric approximation where the electric displacement 
in it is presented as the sum of the constant spontaneous polarization $\pm 
P_{0}$ with antiparallel orientation in neighboring domains and the linear 
dielectric response of the crystal lattice to the electric field, characterized 
by the permittivity $\varepsilon$.

We consider a planar 180${}^\circ$ domain wall, which makes angle $\alpha$ with 
the vector of spontaneous polarization. We address the situation where the tilt 
of the domain wall in the clockwise direction yields the appearance of positive 
bound charges of surface density $\sigma_b=2P_0 \sin\alpha$ as it is shown in 
Fig. \ref{fig01a}. The bound charges are completely compensated by free charges 
of surface density $\sigma_f=-\sigma_b$. The bottom electrode is grounded and 
the top electrode is connected to a source of constant voltage $-U$. The 
voltage applied to the electrodes produces an electric field $E_0=U/h$.

\begin{figure}
    \includegraphics[width=85mm]{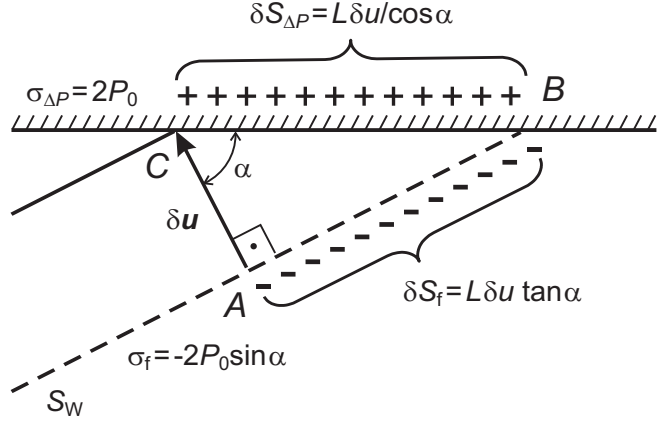}
\caption{Detail of the intersection area of the domain wall and the top
electrode indicating the charge variation on the top electrode during the
virtual displacement of the domain wall. Dashed and solid lines show the
original and displaced positions of the domain wall $S_W$, respectively. Symbol
$\sigma_{\Delta P}$ stands for the surface charge density of bound charges due
to spontaneous polarization reversal during the domain wall shift. Symbol
$\sigma_{f}$ stands for the surface charge density of free charges at the
domain wall. Symbol $\delta S_{\Delta P}$ indicates the area where spontaneous
polarization reversal occurs during the domain wall shift. Symbol $\delta S_f$
indicates the area of domain wall, which reaches the electrode during the
domain wall shift.}
    \label{fig01b}
\end{figure}
In this study, we calculate the value of local pressure on the ferroelectric 
domain wall produced by the electric field using the principle of virtual 
displacements. We consider the wall at rest, when the force imposed on it by 
the electric field is balanced by the external mechanical pressure $p$. 
According to the principle of virtual displacements, the variation of the 
proper thermodynamic function $\delta G$ equals the work produced by the 
external mechanical pressure $p$ during the virtual displacement, i.e., $\delta 
W_p=-S_Wp\delta u$, where $S_W$ is the area of the domain wall and $\delta u$ 
is the virtual displacement of the domain wall in the normal direction. One 
readily checks that for the considered model in the hard ferroelectric 
approximation, the thermodynamic function of the system consists of the 
electrostatic energy and the energy supplied to the system by external electric 
sources; it reads: 
\begin{equation}
    G = \int_V \frac 12\varepsilon E^2\, dV + UQ,
    \label{eq:01:G}
\end{equation}
where $\varepsilon$ is the permittivity of the ferroelectric and $Q$ is the 
charge on the top electrode. Figure \ref{fig01b} shows in detail the 
intersection area of the domain wall and the top electrode. The considered 
virtual displacement is indicated by the line $AC$, where $A$ indicates a point 
on the domain wall in the original position and $C$ indicates the same point on 
the displaced domain wall. Since the bound charges on the oblique domain wall 
are considered to be fully compensated by free charges, the electric field in 
the ferroelectric capacitor is homogeneous and equal to $E_0=U/h$.
This means 
that the variation of the thermodynamic function, $\delta G$, which is produced 
by the virtual displacement $\delta u$ of domain wall, is given only by the 
variation of charge on the top electrode, $\delta Q$. Thus, we can write the 
following equation of equilibrium:
\begin{equation}
    U\, \delta Q = -S_W p\, \delta u.
    \label{eq:02:deltaG}
\end{equation}

Considering the ferroelectric capacitor of length $L$ in the direction
perpendicular to the cross-section shown in Fig. \ref{fig01a}, the area of
oblique domain wall equals
\begin{equation}
    S_W = Lh/\cos\alpha.
    \label{eq:03:SW}
\end{equation}
The variation of charge on the top electrode $\delta Q$ includes the charge 
$\delta Q_{\Delta P}$ due to the spontaneous polarization reversal, which is 
partially compensated by the free charge $\delta Q_f$ that has arrived at the 
top electrode from the ``annihilated" part of the wall during its virtual 
displacement $\delta u$ in the normal direction. Thus it is $\delta Q=\delta 
Q_{\Delta P} + \delta Q_f$. The charge $\delta Q_{\Delta P}$ is provided by the 
bound charge density $\sigma_{\Delta P} = 2P_0$ due to spontaneous polarization 
reversal, which occurs on the area $\delta S_{\Delta P}$ on the top electrode 
indicated by the line $BC$ in Fig. \ref{fig01b}. Using trigonometric functions 
one can readily get the relation between the virtual displacement of the domain 
wall $\delta u$ and the area $\delta S_{\Delta P} = L \delta u/\cos\alpha$. 
Thus the spontaneous polarization reversal produces the variation of charge on 
the top electrode $\delta Q_{\Delta P} = -\sigma_{\Delta P}\delta S_{\Delta P} 
= -2P_0L\delta u/\cos\alpha$. The charge $\delta Q_f$ equals the free charge 
from the piece of the domain wall that ``annihilates" at the electrode during 
the virtual displacement of the domain wall. This charge has a surface density 
of $\sigma_f = -2P_0\sin\alpha$ and is distributed on the area $\delta S_f$ on 
the domain wall indicated by the line $AB$ in Fig. \ref{fig01b}. Using 
trigonometric functions, it is easy to express the relation between the virtual 
domain wall displacement $\delta u$ and the area $\delta S_f = -\sigma_f\delta 
S_f = 2LP_0 \delta u (\sin\alpha)/(\tan\alpha)$. Finally, one can present the 
variation of the charge on the top electrode in the form:
\begin{equation}
    \delta Q = -2P_0L\,\left(1-\sin^2\alpha\right)\,\delta u/\cos\alpha.
    \label{eq:04:deltaQ}
\end{equation}

\begin{figure}
    \includegraphics[width=85mm]{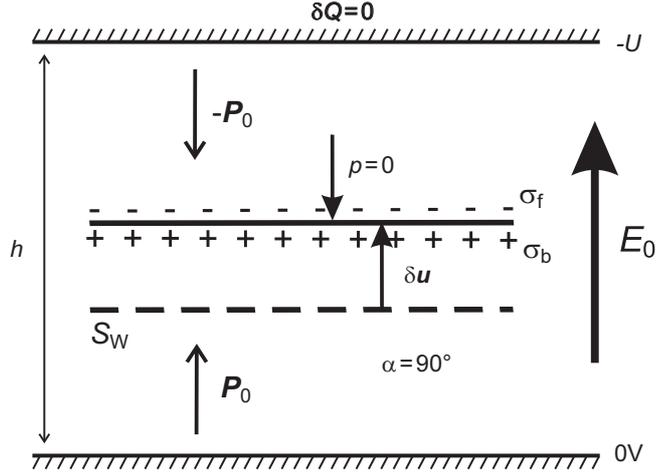}
\caption{Special case of a fully compensated domain wall, which is parallel to 
the electrodes (i.e. $\alpha=90^\circ$). In this case, shift $\delta u$ of the 
domain wall $S_W$ in the normal direction does not produce any change of charge 
on the top electrode, $\delta Q=0$, resulting in zero local pressure on such a 
domain wall.}
    \label{fig01c}
\end{figure}
Substituting Eqs. (\ref{eq:03:SW}) and (\ref{eq:04:deltaQ}) into 
(\ref{eq:02:deltaG}) we get the formula for the local mechanical pressure on an 
inclined 180${}^\circ$ domain wall, which is fully compensated with free 
charges:
\begin{equation}
    p = 2P_0E_0\, (1 - \sin^2\alpha).
    \label{eq:05:p}
\end{equation}
It is seen that, for the domain wall that is normal to the electrode, i.e. 
$\alpha=0$, the above formula reduces to the well-known value of pressure 
$2P_0E$ on an uncharged 180${}^\circ$ domain wall. Figure \ref{fig01c} shows 
the opposite extreme case of a fully compensated domain wall, which is parallel 
to the electrodes (i.e. $\alpha=90^\circ$). This case corresponds to a fully 
compensated head-to-head configuration of the spontaneous polarization. In this 
situation, according to Eq. (\ref{eq:05:p}) the local mechanical pressure on 
the domain wall is zero. This can be readily understood. Indeed, in this case, 
the shift of domain wall $\delta u$ does not produce any change in the 
thermodynamic function $G$, since a displacement of the fully compensated wall 
should affect neither the electric field nor the charge on the top 
electrode.

\section{Local pressure on a ferroelectric domain wall}
\label{sec:Generalized}

In this section we obtain a formula for the local pressure on the ferroelectric 
domain wall for the most general situation where (i) the electric field is 
inhomogeneous, (ii) wall configuration is arbitrary, (iii) charge compensation 
of the wall is not assumed to be full, and (iv) the analysis goes beyond the 
hard ferroelectric approximation. The only limitation of the presented analysis 
would be the neglect of the mechanical effects, i.e. we set the mechanical 
stress $\tau_{ij}^{\rm mech}$ zero everywhere in the sample. Whenever such 
effects are essential they may be incorporated in the framework.

The desired formula can be obtained by application of the virtual displacement 
method to a general configuration of electrodes and arbitrary compensated 
ferroelectric domain walls (see Appendix \ref{sec:Appendix}). However, the 
calculations can be essentially simplified by using an advanced thermodynamic 
result derived in the classical textbook by Landau and Lifshitz \cite{Landau}. 
According to this book, the volume force density $f_i$ in a dielectric can be 
presented in the form
\begin{equation}
    f_i = \frac{\partial\tau_{ij}}{\partial x_j}
    \label{eq:09:volumeforce} 
\end{equation}
in term of the generalized stress tensor $\tau_{ij}$ which, in turn, can be 
written as
\begin{equation}
    \tau_{ij} = \widetilde{\Phi}\,\delta_{ij} + \tau_{ij}^{\rm mech} + \frac 12\, 
            \left(E_iD_j + E_jD_i\right)
    \label{eq:10:genforce} 
\end{equation}
with
\begin{equation}
    \widetilde{\Phi} = \Phi - (1/2)\, \varepsilon_0E^2 - E_i P_i
    \label{eq:08:bulkenergy} 
\end{equation}
where $P_i$ and $D_i=P_i+\varepsilon_0\,E_i$ are the polarization and 
electrical displacement, respectively; $\Phi$ is the thermodynamic function of 
the dielectric characterized by the differential $d\Phi=E_i\,dP_i + 
\tau_{ij}^{\rm mech}\,du_{ij}$ where $u_{ij}$ is the strain tensor.

Applying Eq. (\ref{eq:09:volumeforce}) to the two domains in the vicinity of a 
domain wall, one presents the pressure to the latter $p$ in the form
\begin{equation}
    p=\DifrOp{\tau_{ij}}\, n_i n_j 
    \label{eq:12:pdefined}
\end{equation}
where $n_i$ is vector normal to the wall. The notation $\DifrOp{Z}$ here and 
therein is used to denote the jump of the variable $Z$ at the wall. Considering 
the mechanically free ferroelectric sample, i.e. $\tau_{ij}^{\rm mech}=0$, and 
combining Eqs. (\ref{eq:10:genforce}) and (\ref{eq:12:pdefined}) one gets
\begin{equation}
    p=[[\widetilde{\Phi}]] + \DifrOp{E_iD_j}\, n_i n_j.
    \label{eq:13:pisgen}
\end{equation}
If we further employ the expressions for the continuity of tangential 
components of the electric field at the domain wall $\DifrOp{E_{t,i}} = 
\DifrOp{E_i - (E_kn_k) n_i} = 0$, the relation for the normal component of 
electric displacement at the domain wall $\DifrOp{D_i}\, n_i = \sigma_f$, and 
the algebraic identity
\begin{equation}
    \label{eq:14:AlgebrId}
    \DifrOp{E_iD_j} = \widehat{E}_i\DifrOp{D_j} + \DifrOp{E_i}\widehat{D}_j,
\end{equation}
where $\widehat{E}_i = (\1{E_i} + \2{E_i})/2$ is the average electric field at 
the opposite sides of the domain wall, we get the formula for the external 
mechanical pressure on the domain wall
\begin{equation}
    \label{eq:15:pRes}
    p = [[\widetilde{\Phi} + (1/2)\, \varepsilon_0E^2 + P_iE_i]]
        - \widehat{E}_i\left(\DifrOp{P_i} -
        \sigma_f n_i\right).
\end{equation}
Considering Eq. (\ref{eq:08:bulkenergy}), formula (\ref{eq:15:pRes}) can be 
further simplified:
\begin{eqnarray}
    \label{eq:16:pIs}
    p = \Phi^{(2)} - \Phi^{(1)}
        - \frac 12\left(E_i^{(1)} + E_i^{(2)}\right) \left(P_i^{(2)} - P_i^{(1)} -
        \sigma_f n_i\right).
\end{eqnarray}
Here upper indexes $(1)$ and $(2)$ specify to which domain the variable is 
referred to, the direction of the normal vector $n_i$ is taken inside domain 
$(2)$, and the pressure is considered as positive when acting from domain $(2)$ 
to domain $(1)$.

\section{Pressure on the wall in special cases}
\label{sec:PressureSpecial}

In this section, we present consequences of Eq. (\ref{eq:16:pIs}) and 
demonstrate its applicability for two particular cases. At first we present 
further a simplification of formula (\ref{eq:16:pIs}) for the case of low 
applied fields where the dielectric non-linearity of the ferroelectric can be 
neglected. Then we demonstrate the application of formula (\ref{eq:16:pIs}) in 
systems where the non-linearity of the dielectric response of the crystal 
lattice is essential.

\subsection{Hard ferroelectric approximation}
\label{sec:sub:LowFields}

For small electric fields we can use the ``hard ferroelectric" approximation 
and express the polarization as a sum of the constant spontaneous polarization 
$P_{0,i}$ and the linear response of the polarization of the crystal lattice to 
the electric field
\begin{subequations}
\label{eq:18:PSmall}
\begin{equation}
    \label{eq:18a:PSmallField}
    P_i \approx P_{0,i} + \chi_{ij} E_j,
\end{equation}
where $\chi_{ij}$ is the susceptibility of the ferroelectric. In this situation,
function $\Phi(P)$ can be expressed in a form
\begin{equation}
    \label{eq:18b:PhiSmallField}
    \Phi(P) \approx \frac 12 \chi_{ij} E_iE_j.
\end{equation}
\end{subequations}
After substitution of Eqs. (\ref{eq:18:PSmall}) into Eq. (\ref{eq:16:pIs}) one 
readily gets
\begin{equation}
    \label{eq:19:pSmallField}
    p = -\widehat{E}_i\left(\DifrOp{P_{0,i}} -
        \sigma_f n_i\right) -
        \frac 12 \DifrOp{\chi_{ij}} \1{E_i}\2{E_j}.
\end{equation}
In the case of absence of free charges in the system, Eq. 
(\ref{eq:19:pSmallField}) was obtained earlier\cite{Nechaev,Kessler}.

\begin{figure}
    \includegraphics[width=85mm]{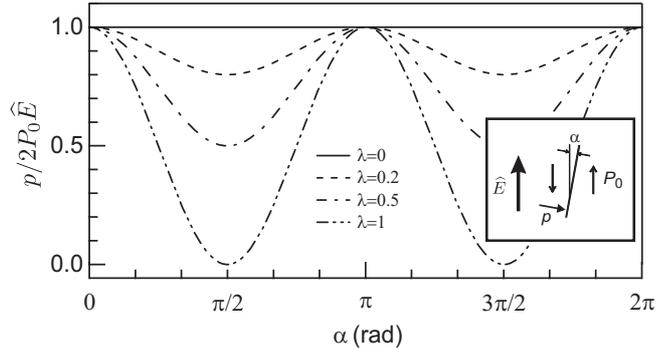}
    \caption{Normalized pressure on 180${}^\circ$ ferroelectric domain wall
versus the tilt of domain wall $\alpha$ with respect to vectors of
spontaneous polarization, which is shown in the inset, for
different degrees of compensation of bound charge $\sigma_b$ by free
charge $\sigma_f$; $\lambda=-\sigma_f/\sigma_b$.}
    \label{fig03}
\end{figure}
\begin{figure}
    \includegraphics[width=85mm]{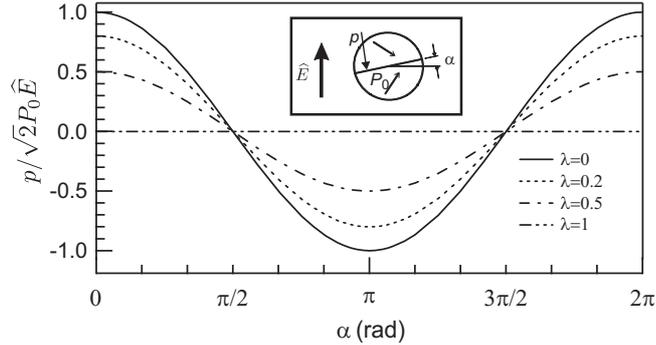}
\caption{Normalized pressure on a 90${}^\circ$ ferroelectric domain wall versus 
the angle, $\alpha$, between the normal to the wall and the vector of electric 
field, which is shown in the inset, for different degrees of compensation of 
bound charge $\sigma_b$ by free charges $\sigma_f$; 
$\lambda=-\sigma_f/\sigma_b$. Note that for complete screening of bound charges,
 i.e. $\lambda=1$, the local pressure on the domain wall is zero for any 
relative orientation of the domain wall and the electric field. Here the 
contribution due to the dielectric anisotropy is neglected, i.e. 
$\DifrOp{\chi_{ij}}=0$}
    \label{fig04}
\end{figure}

Two numerical examples of the effect of free charges on the local pressure on a 
charged ferroelectric domain wall are shown in Figs. \ref{fig03} and 
\ref{fig04}. In Fig. \ref{fig03} we consider the case of a 180${}^\circ$ 
ferroelectric domain wall where the vector of average electric field 
$\widehat{E}_i$ is considered to be always parallel to the vectors of 
spontaneous polarization $P_{0,i}$ and the domain wall is considered to be 
tilted by an angle $\alpha$ with respect to the directions of spontaneous 
polarization. The tilt of the domain wall results in the appearance of bound 
charge on the domain wall, $\sigma_b$, and the local pressure of external 
sources on such a wall equals $2P_{0,i}\widehat{E}_i$ and its direction is 
shown in the inset of Fig.\ref{fig03}. Then we consider that the bound charge 
is partially compensated by free charges $\sigma_f$. The degree of partial 
compensation is expressed by the quantity $\lambda$, so that 
$\sigma_f=-\lambda\sigma_b$ where $0\le\lambda\le 1$. It is seen that free 
charges reduces the value of pressure $p$, but the important point is that, for 
the same degree of screening $\lambda$, the value of $p$ varies depending on 
the domain wall tilt. Zero pressure corresponds to the situation of full 
compensation of the domain wall, which is perpendicular to the vectors of 
spontaneous polarization.

The effect of free charge compensation on the pressure on a charged 
90${}^\circ$ domain wall is shown in Fig. \ref{fig04}. However, the 
contribution due to dielectric anisotropy is neglected (i.e. 
$\DifrOp{\chi_{ij}}=0$), since it produces a field redistribution\cite{Novak}, 
which can be hardly evaluated in the general case. Since this type of domain 
wall is ferroelastic it is reasonable to consider that the angle between the 
vectors of spontaneous polarization and the domain wall is fixed as it is 
indicated in the inset of Fig. \ref{fig04}. The domain wall charge is equal to 
$\sigma_b=\sqrt{2}P_{0}$ and the local pressure is a function of the tilt of 
the domain wall with respect to the average electric field $\widehat{E}_i$. It 
is seen that, in this case, the presence of free charges leads to an effective 
decrease of the value of spontaneous polarization. The physical reason for this 
is that in this case the vector $\DifrOp{P_{0,i}}$ is always parallel to the 
normal vector of the domain wall $n_i$. It is noticeable that, in the case of 
the complete bound charge compensation, the pressure on the domain wall is zero 
regardless the orientation of the average electric field and the domain 
wall.

\subsection{Nonlinear effects}
\label{sec:sub:NonlinearEffects}

Formula (\ref{eq:16:pIs}) makes it possible to analyze the effects of the 
dielectric nonlinearity of the crystal lattice on the local pressure $p$. As an 
example we consider a fully compensated 180${}^\circ$ domain wall, which is 
perpendicular to the vectors of spontaneous polarization (see Fig. \ref{fig01b})
. In section \ref{sec:Concept} we have shown that, in the hard ferroelectric 
approximation, the local pressure on such a wall is zero. However, analysis 
beyond this approximation shows that this force is not actually zero. Let us 
find this force for the case of a uniaxial ferroelectric with the second order 
phase transition. In this case, the free energy of the ferroelectric reads:
\begin{equation}
    \label{eq:20:FreeEnHOTerms}
    \Phi(P) = \Phi_0 + \frac 12\alpha P^2 + \frac 14\beta P^4,
\end{equation}
where, in the ferroelectric phase, $\alpha<0$ and $\beta>0$. Discontinuous 
change of polarization states $\1{P}$ and $\2{P}$ at the domain wall and the 
presence of free charges $\sigma_f$ produce the electric field $\Delta E$ 
according to the equation:
\begin{subequations}
\label{eq:21:NonlinearEq}
\begin{equation}
    \2{P} - \1{P} - \sigma_f = 2\varepsilon_0 \Delta E.
\end{equation}
This electric field is of antiparallel orientation at the opposite sides of the 
domain wall and it is superposed on the uniform applied electric field $E_0$, 
so that the equation of state in the adjacent domains are of a form:
\begin{eqnarray}
    \partial \Phi(\1{P})/\partial P &=& E_0 + \Delta E, \\  
    \partial \Phi(\2{P})/\partial P &=& E_0 - \Delta E.     
\end{eqnarray}  
\end{subequations}
The polarization $\1{P}$ and $\2{P}$ at the opposite sides of the domain wall 
is given by the solution of Eqs. (\ref{eq:21:NonlinearEq}) and, in the case of 
fully compensated domain wall, i.e., $\sigma_f=-2P_0$, it can be expanded in a 
Taylor series with respect to the applied electric field $E_0$:
\begin{subequations}
\label{eq:22:P}
\begin{eqnarray}
    \label{eq:22a:P1}
    \1{P}&\approx& P_0 + \chi E_0 - \frac {3\varepsilon_0\chi^2}{2P_0 (\chi+\varepsilon_0)}E_0^2 + \cdots,\\
    \label{eq:22b:P2}
    \2{P}&\approx& -P_0 + \chi E_0 + \frac {3\varepsilon_0\chi^2}{2P_0 (\chi+\varepsilon_0)}E_0^2 + \cdots,
\end{eqnarray}
\end{subequations}
where $P_0=\sqrt{-\beta/\alpha}$ is the spontaneous polarization and 
$\chi=-1/(2\alpha)$ is the dielectric susceptibility of the crystal lattice. 
After substitution of Eqs. (\ref{eq:22:P}) into Eq. (\ref{eq:20:FreeEnHOTerms}),
 we can readily obtain the expansion of function $\Phi(P)$ at the opposite 
sides of the domain wall in terms of the applied electric field:
\begin{subequations}
\label{eq:23:Phi}
\begin{eqnarray}
    \label{eq:23a:Phi1}
    \Phi(\1{P})&\approx& -\frac{P_0^2}{8\chi} + \frac 12\chi E_0^2 + \frac {\chi^2(\chi-2\varepsilon_0)}{2P_0(\chi+\varepsilon_0)}E_0^3 + \cdots,\\
    \label{eq:23b:Phi2}
    \Phi(\2{P})&\approx& -\frac{P_0^2}{8\chi} + \frac 12\chi E_0^2 - \frac {\chi^2(\chi-2\varepsilon_0)}{2P_0(\chi+\varepsilon_0)}E_0^3 + \cdots.
\end{eqnarray}
\end{subequations}
When one considers that the average field in the ferroelectric,
$\widehat{E}$, is dominated by the applied electric field $E_0$, i.e.
$\widehat{E}=E_0$, the resulting formula for the pressure on the domain wall is 
according to Eq. (\ref{eq:16:pIs}) equal to:
\begin{equation}
    \label{eq:24:p}
    p\approx -\left(\chi^2/P_0\right)\, E_0^3,
\end{equation}
which means that the pressure on the domain wall is oriented in the opposite 
direction than the ``expected" $2P_0E_0$ contribution.

\begin{figure}
    \includegraphics[width=85mm]{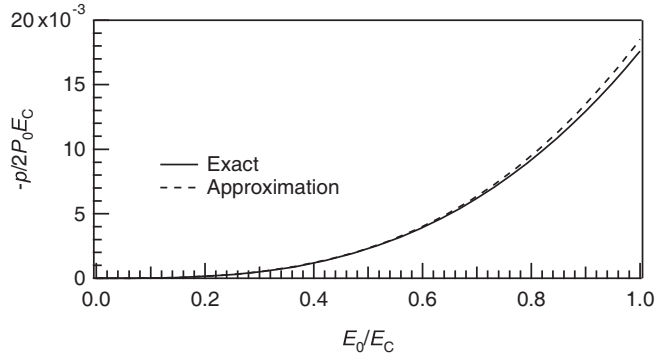}
\caption{
The applied field dependence of the local pressure on a fully compensated 
head-to-head 180$^{\circ}$ ferroelectric domain wall. The pressure is 
normalized to $-2P_0E_C$, that equals the value of pressure expected to be 
applied to an uncompensated domain wall at the thermodynamic coercive field 
$E_C$ in the hard ferroelectric approximation. The field is normalized to the 
thermodynamic coercive field $E_C$. The exact values (solid line) obtained by 
numerical solving the nonlinear Eqs. (\ref{eq:21:NonlinearEq}) are compared 
with the approximative values (dashed line) calculated using the formula 
(\ref{eq:24:p}).
    }
    \label{fig05}
\end{figure}
The physical interpretation of the above formula is as follows. The application 
of a high electric field to the ferroelectric sample with the considered 
configuration of a fully compensated charged 180${}^\circ$ domain wall is 
associated with two effects: first, it produces the difference in volume energy 
density of the crystal lattice at the opposite sides of the domain wall due to 
dielectric nonlinearity of the crystal lattice, and second, it produces an 
additional charging of the domain wall due to a different polarization response 
at opposite sides of the domain wall so that the wall becomes negatively 
charged. The pressure associated with the electrostatic force acting on this 
charge is directed against that associated with the energy difference. Finally, 
the resulting formula corresponds to the sum of these effects.

Figure \ref{fig05} shows the applied electric field dependence of the local 
pressure on a fully compensated domain wall, which is parallel to the 
electrodes (see Fig. \ref{fig01c}). The dashed line shows the approximative 
result given by formula (\ref{eq:24:p}). The solid line shows the exact result 
given by the numeric solution of the system of nonlinear equations 
(\ref{eq:21:NonlinearEq}), where the function $\Phi(P)$ is given by Eq. 
(\ref{eq:20:FreeEnHOTerms}). In Fig. \ref{fig05}, numerical values typical for 
barium titanate were used: $P_0=0.26\,$Cm${}^{-2}$ and $\chi=188\, 
\varepsilon_0$. It is seen that Eq. (\ref{eq:24:p}) provides a very good 
approximation, which differs only by about 6\% from the exact value at the 
thermodynamic coercive field $E_C=P_0/\left(3\sqrt{3}\chi\right)$.

\section{Additional imprint mechanism}

The results obtained in the previous sections enable us to draw a conclusion on 
the imprint behavior of polydomain ferroelectrics.

Imprint is a property of a ferroelectric capacitor of exhibiting a higher 
switching voltage when the switching starts from a state in which the capacitor 
has been kept for a long time or subjected to light illumination (see Ref. 
\cite{Tagantsev} for comprehensive discussion). A popular and widely referred 
to imprint mechanism, which has been offered by the Sandia Laboratory group 
\cite{Warren1,Warren2}, is related to free-carrier compensation of charged 
domain walls. Our analysis presented in this paper suggests that this 
compensation should lead to one more imprint mechanism.

The key issue of the Sandia mechanism, which we will refer to as ``trapping 
imprint", is that, if the free carriers compensating the bound charge of a wall 
are trapped, then the mobility of this wall is deteriorated since now it is to 
drag with it the weakly mobile trapped carries. Since, in typically 
low-conductive ferroelectrics, an appreciable time or/and illumination is 
needed for the free-carrier compensation of the bound charge of a wall, this 
effect will lead to the imprint. In the context of the analysis presented above 
its is clear that the aforementioned charge compensation will also lead to a 
reduction of the ``thermodynamic" (pondermotoric) force applied to the wall in 
the switching field. This reduction, similarly to the above case, can manifest 
itself in an imprint mechanism, which we will refer to as ``force-reduction" 
imprint. One should stress that the performance of these mechanisms can be 
appreciably different. For instance, in the case of trapping imprint, if the 
switching field is applied long enough to let the trapped charges drift with 
the wall, the imprint effect becomes weaker. Its efficiency is clearly 
controlled by the mobility of the trapped free carriers; in the limit of high 
mobility the effect vanishes. On the other hand, the force-reduction imprint is 
clearly insensitive to the mobility of the compensating free chargers. In 
contrast to the trapping imprint, one cannot cope with this mechanism by using 
longer switching pulses. All in all, one sees a possibility to distinguish 
these imprint mechanisms in real ferroelectric systems.

\section{Conclusions}

The impact of free charges on the local pressure on a charged ferroelectric 
domain wall produced by an electric field has been analyzed. A general formula 
for the local pressure on a charged domain wall is derived considering full or 
partial compensation of the bound polarization charge with the free charge. It 
is shown that the compensation can lead to a very strong reduction of the 
pressure imposed on the wall by the electric field. In some cases this pressure 
can be governed by small nonlinear effects. It is concluded that the free 
charge compensation of the bound polarization charge can lead to substantial 
reduction of the domain wall mobility even in the case of the high mobility of 
the free charge. This mobility reduction gives to an additional imprint 
mechanism which may play essential role in switching properties of 
ferroelectric materials. Using this mechanism, it is possible to explain 
frequent observations of very stable domain patterns with the head-to-head 
configuration of the vectors of spontaneous polarization \cite{Rodriguez}. The 
results obtained can be used in modeling the poling of ferroelectric ceramics, 
domain nucleation, sidewise domain wall movement and spontaneous polarization 
switching in imprinted, leaky ferroelectric samples, and samples treated at 
elevated temperatures.

\begin{acknowledgements}

This project was supported by the Swiss National Science Foundation and by the 
Grant Agency of the Czech Republic, project GACR 202/06/0411. Authors 
acknowledge Guido Gerra for reading the manuscript.

\end{acknowledgements}


\appendix

\section{Local pressure on a ferroelectric domain wall derived from the principle of virtual displacements}
\label{sec:Appendix}

\begin{figure}[t]
    \includegraphics[width=85mm]{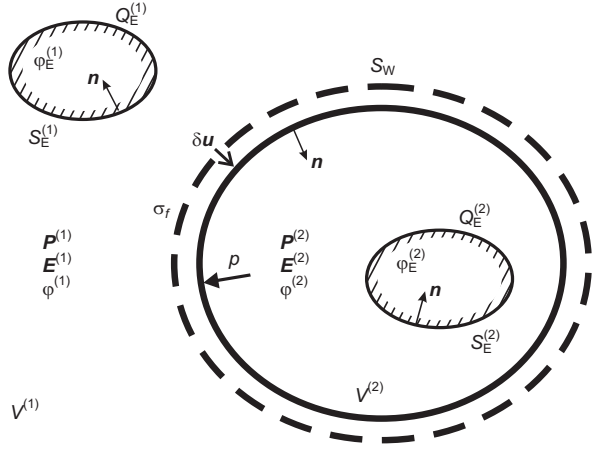}
\caption{Configuration used to calculate the local pressure imposed by the
electric field to a wall. Domain wall $S_W$ separates the ferroelectric into
two domains of volumes $\1{V}$ and $\2{V}$. There are two conductors, which
carry charges $\1{Q_E}$ and $\2{Q_E}$ and have electric potentials
$\1{\varphi_E}$ and $\2{\varphi_E}$ within each domain. The bulk quantities
within each domain are the polarization $\1{P_i}$ and $\2{P_i}$, electric field
$\1{E_i}$ and $\2{E_i}$ and electric potential $\1{\varphi}$ and $\2{\varphi}$.
Symbol $\delta u$ stands for the virtual displacement of the domain wall and
symbol $p$ stands for the external mechanical pressure that should be applied
to the wall to keep it at rest.}
    \label{fig06}
\end{figure}

In this Appendix we offer an alternative derivation of the local pressure on 
ferroelectric domain wall than it is presented in Sec. \ref{sec:Generalized}. 
Here, we use again the principle of virtual displacements as we do in Sec. 
\ref{sec:Concept}. We consider a system shown in Fig. \ref{fig06} where inside 
a ferroelectric material with polarization $\1{P_i}$ there exists a domain of 
the material with a different polarization state $\2{P_i}$, i.e. that there is 
a discontinuity of the polarization across the interface $S_W$. This interface 
splits the ferroelectric into two domains, which have volumes $\1{V}$ and 
$\2{V}$, respectively. Inside each domain there is a conductor, which carries 
charge $\1{Q_E}$ and $\2{Q_E}$, respectively, and has electric potential 
$\1{\varphi_E}$ and $\2{\varphi_E}$, respectively. We consider that bound 
charges due to discontinuous change of polarization at the domain wall $S_W$ 
are partially compensated by free charges of surface density $\sigma_f$. The 
charges on conductors and on the domain wall produce electric fields $\1{E_i}$ 
and $\2{E_i}$ within each domain. Symbols $\1{\varphi}$ and $\2{\varphi}$ stand 
for electric potentials within each domain.

In order to obtain a general formula for the external mechanical pressure $p$ 
on the domain wall, we follow the principle of virtual displacements presented 
in Section \ref{sec:Concept}. However, we adopt the thermodynamic function $G$ 
in a more general form:
\begin{equation}
    G = \int_V \left[\Phi(P) + \frac 12\varepsilon_0 E^2\right]\, dV - \sum_i\varphi_E^{(i)}\, Q_E^{(i)},
    \label{eq:ap:01:G}
\end{equation}
The first term in the above formula represents the energy associated with the
polarization of crystal lattice of ferroelectric, $\Phi(P)$, and the electric
field energy, $(1/2)\varepsilon_0 E^2$, which are integrated over the volume
$V$ of ferroelectric. The second term in Eq. (\ref{eq:ap:01:G}) represents the
subtracted work of the electric sources. It should be also noticed that the
thermodynamic function $G$ given by Eq. (\ref{eq:ap:01:G}) is used just for
convenience; the result (equation of state) is independent of the choice of the
potential as always.

In what follows it is convenient to transform the work of electric
sources into volume integrals:
\begin{subequations}
\label{eq:ap:02:IntIds}
\begin{eqnarray}
    \label{eq:ap:02a:IntId1}
    \varphi_E^{(1)}\, Q_E^{(1)} &=& - \varphi_E^{(1)}\,
    \int_{S_E^{(1)}}D_i^{(1)} n_i\, dS =
    \int_{V^{(1)}} E_i^{(1)} D_i^{(1)}\, dV
        +
        \int_{S_W} \varphi^{(1)} D_i^{(1)} n_i\, dS, \\
    \label{eq:ap:02b:IntId1}
    \varphi_E^{(2)}\, Q_E^{(2)} &=& - \varphi_E^{(2)}\,
    \int_{S_E^{(2)}}D_i^{(2)} n_i\, dS = \int_{V^{(2)}} E_i^{(2)} D_i^{(2)}\, dV
        -
        \int_{S_W} \varphi^{(2)} D_i^{(2)} n_i\, dS,
\end{eqnarray}
\end{subequations}
where $\1{D_i}$ and $\2{D_i}$ are the vectors of electric displacement in 
domains $\1{V}$ and $\2{V}$, respectively. In the derivation of the above 
equations, we consider the absence of free charges, i.e. $\dv D_i=0$, inside 
domains $V^{(1)}$ and $V^{(2)}$ and vanishing the surface integral over the 
domain $V^{(1)}$ at infinity. Orientation of the normal vectors $n_i$ is 
indicated in Fig. \ref{fig06}. If we further employ the expressions for 
electric displacement $D_i=P_i+\varepsilon_0\,E_i$, the continuity of the 
electric potential at the domain wall $\DifrOp{\varphi}=0$, the continuity of 
the normal component of electric displacement at the domain wall $\DifrOp{D_i}\,
 n_i = \sigma_f$, and the algebraic identity used in Eq. (\ref{eq:14:AlgebrId}),
 the thermodynamic function $G$ can be expressed in the form:
\begin{eqnarray}
    \label{eq:ap:05:G}
    G &=& \int_{\1{V}}\left[
        \1{\Phi}(\1{P}) - \frac 12 \varepsilon_0\1{E_i}\1{E_i}
        - \1{E_i}\1{P_i}
    \right]\, dV + \\
    \nonumber
    &+& \int_{\2{V}}\left[
        \2{\Phi}(\2{P}) - \frac 12 \varepsilon_0\2{E_i}\2{E_i}
        - \2{E_i}\2{P_i}
    \right]\, dV + \\
    &+& \int_{S_W}\widehat{\varphi}\,\sigma_f\, dS.
    \nonumber
\end{eqnarray}
For the sake of presentation, it will be useful to define functions
\begin{subequations}
\begin{eqnarray}
    \widetilde{\Phi}_B &=& \Phi(P) - \frac 12 \varepsilon_0E_iE_i - E_i P_i,
    \label{eq:ap:06a:bulkenergy} \\
    \widetilde{\Phi}_S &=& \widehat{\varphi}\,\sigma_f
    \label{eq:ap:06b:surfaceenergy}
\end{eqnarray}
\end{subequations}
and to write the function $G$ in the form
\begin{equation}
    G = \int_{V} \widetilde{\Phi}_B\,dV + \int_{S_W} \widetilde{\Phi}_S\, dS,
    \label{eq:ap:07:Ggeneral}
\end{equation}
where the volume integral is taken over the volume $V=\1{V}+\2{V}$.

Now our task is to express the variation of thermodynamic function $\delta G$, 
which is produced by the virtual displacement of domain wall:
\begin{equation}
    \delta G = \int_{V} \delta \widetilde{\Phi}_B\,dV + \int_{S_W} \delta \widetilde{\Phi}_S\,
    dS - \int_{S_W} \left\{ \DifrOp{\widetilde{\Phi}_B} - \frac{\partial \widetilde{\Phi}_S}{\partial x_k}\,n_k\right\}\, \delta u\,
    dS,
    \label{eq:ap:08:varGgeneral}
\end{equation}
where the first two terms represent the variations $\delta \widetilde{\Phi}_B$ 
and $\delta \widetilde{\Phi}_S$ due to a change in bulk quantities during the 
virtual displacement of domain wall $\delta u$. The last term represents the 
contribution to the variation $\delta G$ due to the volume change of domains 
produced by the virtual displacement of domain wall $\delta u$ and due to the 
change of the position of the domain wall. At equilibrium, the variation 
$\delta G$ equals the work $\delta W_p$ produced by external mechanical 
pressure $p$ during the virtual displacement of the domain wall
\begin{equation}
    \label{eq:ap:09:deltaWp}
    \delta W_p= - \int_{S_W}p\delta u\, dS.
\end{equation}
Employing the principle of virtual displacements $\delta G=\delta W_p\,$, it 
can be readily shown that the first two terms in Eq. 
(\ref{eq:ap:08:varGgeneral}) vanish because of the equations of state in the 
bulk ferroelectric
\begin{equation}
    \label{eq:ap:10:EqOfMotion}
    \frac{\partial\Phi}{\partial P_i} = E_i,
\end{equation}
Gauss' law for electric displacement $\partial D_i/\partial x_i = 0$, and the 
continuity of the normal component of electric displacement at the domain wall 
$\DifrOp{D_i} n_i = \sigma_f$. Finally, the last term in Eq. 
(\ref{eq:ap:08:varGgeneral}) yield the formula for the external mechanical 
pressure on the domain wall
\begin{equation}
    \label{eq:ap:11:pIs}
    p = \DifrOp{
                \Phi(P) -
                \frac 12 \varepsilon_0 E_i E_i -
                E_i P_i
            } -
            \widehat{E_i}\, \sigma_f\, n_i.
\end{equation}
Taking into account the continuity of tangential components of the electric 
field at the domain wall, $\DifrOp{E_{t,i}} = \DifrOp{E_i - (E_kn_k) n_i} = 0$, 
continuity of the electrostatic potential at the domain wall 
$\DifrOp{\varphi}=0$, and the algebraic identity (\ref{eq:14:AlgebrId}), we 
thus arrive at the result announced above, Eq. (\ref{eq:16:pIs}), i.e.:
\begin{equation}
    \label{eq:ap:12:pIs}
    p = \DifrOp{\Phi(P)}
        - \widehat{E}_i\left(\DifrOp{P_i} -
        \sigma_f n_i\right).
\end{equation}

\end{document}